\documentclass[12pt,preprint]{aastex}

\shorttitle{The Chromospheric Activity, Age, Metallicity and Space Motions of 27 Wide Binaries}
\shortauthors{Zhao et al.}

\begin{document}

%% LaTeX will automatically break titles if they run longer than
%% one line. However, you may use \\ to force a line break if
%% you desire.

\title{The Chromospheric Activity, Age, Metallicty and Space Motions of 36 Wide Binaries }

%% Use \author, \affil, and the \and command to format
%% author and affiliation information.
%% Note that \email has replaced the old \authoremail command
%% from AASTeX v4.0. You can use \email to mark an email address
%% anywhere in the paper, not just in the front matter.
%% As in the title, use \\ to force line breaks.

\author{J. K. Zhao\altaffilmark{1,2}, T. D. Oswalt\altaffilmark{1}, M. Rudkin\altaffilmark{1}, G. Zhao\altaffilmark{2}, Y. Q. Chen\altaffilmark{2}}
%\affil{Florida Institute of Technology, Melbourne, US 32901}
\email{jzhao@fit.edu}
\email{toswalt@fit.edu}
\email{mrudkin@fit.edu}
\email{gzhao@bao.ac.cn}
\email{cyq@bao.ac.cn}

%% Notice that each of these authors has alternate affiliations, which
%% are identified by the \altaffilmark after each name.  Specify alternate
%% affiliation information with \altaffiltext, with one command per each
%% affiliation.

\altaffiltext{1}{Florida Institute of Technology, Melbourne, USA, 32901}
\altaffiltext{2}{Key Laboratory of Optical Astronomy, National Astronomical Observatories, Chinese Academy of Sciences, Beijing, 100012, China}

%% Mark off your abstract in the ``abstract'' environment. In the manuscript
%% style, abstract will output a Received/Accepted line after the
%% title and affiliation information. No date will appear since the author
%% does not have this information. The dates will be filled in by the
%% editorial office after submission.

\begin{abstract}
We present the chromospheric activity (CA) levels, metallicities and full space motions for 41 F, G, K and M dwarf stars in 36 wide binary systems. Thirty-one of the binaries, contain a white dwarf component. In such binaries the total
age can be estimated by adding the cooling age of the white dwarf to an estimate of the
progenitor's main sequence lifetime. To better understand how CA correlates to stellar age, 14 cluster member stars were also observed. Our observations demonstrate for the first time that in general CA decays with age from 50 Myr to at least 8 Gyr for stars with 1.0 $\leq$ V-I $\leq$ 2.4. However, little change occurs in CA level for stars with V-I $<$ 1.0  between 1 Gyr and 5 Gyr, consistent with the results of Pace et al. (2009). Our sample also exhibits a negative correlation between stellar age and metallicity, a positive correlation between stellar age
and W space velocity component and the W velocity dispersion increases with age.  Finally, the population membership of these wide binaries is examined
based upon their U, V, W kinematics, metallicity and CA. We conclude that wide binaries are similar to field and cluster stars in these respects. More importantly, they span a much more continuous range in age and metallicity than is afforded by nearby clusters.

%This is partly consistent with the conclusion of Pace $\&$ Pasquini (2009) who suggested
%contradicting the conclusion of Pace $\&$ Pasquini (2009) who suggested
%that the chromospheric activity's decay stops after about 1.5 Gyr.
%% what keyword punctuation is appropriate.
\end{abstract}

%% From the front matter, we move on to the body of the paper.
%% In the first two sections, notice the use of the natbib \citep
%% and \citet commands to identify citations.  The citations are
%% tied to the reference list via symbolic KEYs. The KEY corresponds
%% to the KEY in the \bibitem in the reference list below. We have
%% chosen the first three characters of the first author's name plus
%% the last two numeral of the year of publication as our KEY for
%% each reference.

%% Authors who wish to have the most important objects in their paper
%% linked in the electronic edition to a data center may do so by tagging
%% their objects with \objectname{} or \object{}.  Each macro takes the
%% object name as its required argument. The optional, square-bracket
%% argument should be used in cases where the data center identification
%% differs from what is to be printed in the paper.  The text appearing
%% in curly braces is what will appear in print in the published paper.
%% If the object name is recognized by the data centers, it will be linked
%% in the electronic edition to the object data available at the data centers
%%
%% Note that for sources with brackets in their names, e.g. [WEG2004] 14h-090,
%% the brackets must be escaped with backslashes when used in the first
%% square-bracket argument, for instance, \object[\[WEG2004\] 14h-090]{90}).
%%  Otherwise, LaTeX will issue an error.

\section{Introduction}
Age is one of the most difficult to determine properties of a star.  The Vogt-Russell theorem asserts that the structure of a star is uniquely determined by its mass and composition.  Nucleosynthesis in the core results in changes in composition and this implies at least some measurable property(ies) of a star must vary with age.  Unfortunately, these changes are subtle and difficult to measure.  It is ironic that the age of the Universe (13.7 $\pm$ 0.2 Gyr; Bennett et al. 2003) is known to better precision than the age of any star other than the Sun.  The present methods by which stellar ages can be estimated are seldom consistent within 50\% (Soderblom 2010). Even the Sun does not reveal its age directly; this key calibration point is determined from the decay of radioisotope samples to be 4,566$_{-1}^{+2}$ Myr (Chaussidon 2007).

	One of the `semifundamental' methods of stellar age determination is isochrone fitting the position of a star in the Hertzsprung-Russell diagram (HRD).  However, because of the degeneracy of theoretical isochrones, this technique does not work well for the vast majority of stars-those on the lower main sequence (MS).  Here small errors in luminosity or metallicity translate into large errors in age.

CaII H$\&$K features in the violet spectra of MS stars are one of the more well-studied indicators of CA. Early work by Wilson (1963; 1968) and
 Vaughan $\&$ Preston (1980) established CaII H$\&$K emission as a useful marker of CA in lower MS
 stars. In F to early M stars Skumanich (1972) found that CaII H$\&$K emission, magnetic
  field strength and rotation all decay as the inverse square root of stellar age.

 Mamajek $\&$ Hillenbrand (2008) and others have shown that the CA vs. age relation is much more complex than Skumanich envisioned; such factors as metallicity, photospheric contamination of CA indices and variation in CA must be considered.  Clusters provide only a limited range of ages and metallicities to investigate these effects.

A self-sustaining magnetic dynamo driven by rotation
and convection is believed to be the source of CA in  MS stars of spectral type F, G, K and early M. According to this paradigm, due to magnetic breaking, rotational velocity decreases with
age, which leads to a decrease in CA as well, unless  angular momentum is sustained
by tidal interaction, as in the case of short-period binaries, or maintained by convection as in late M type dwarfs.

 Using a lower resolution analog of the Mount Wilson S index in eight clusters and the Sun, Barry, Cromwell $\&$ Hege (1987) found that the decay of CA is well-represented by an exponential over the age range from $10^{7}$ to 6 $\times$ $10^{9}$ yr.  Barry (1988) adjusted the age of three clusters from Barry, Cromwell $\&$ Hege (1987). Using a color correction C$_{cf}'$ from Noyes et al. (1984), Barry found empirically that CA $\sim$ t$^{1/e}$. In addition, he concluded that the star formation rate in the solar neighborhood has not been constant, suggesting a recent burst of star formation because of the large number of stars in his youngest age bins.  Working toward a more detailed understanding of the CA vs. age relation, Soderblom, Duncan $\&$ Johnson (1991) found that, while a power law is generally the best fit
to the CaII H$\&$K vs. age relation, it implies a constant star formation rate (SFR). Any different SFR causes the Skumanich relation to indicate
an excess of young stars in the solar neighborhood.  They also found that the calibrated cluster data presented in Barry, Cromwell $\&$ Hege (1987) were consistent with a constant SFR.

Pace $\&$ Pasquini (2004) found that for several clusters
older than 1 Gyr there appeared to be a constant activity level. Pace et al. (2009) believed stars change from active to inactive, crossing the activity range corresponding to the so-called `Vaughan-Preston gap', on a time-scale that might be as short as 200 Myr. If true, this would bring into question whether the Skumanich relation is valid for MS stars in all age regimes.

 Among wide white dwarf (WD)+dM binaries Silvestri, Hawley $\&$ Oswalt (2005) used the cooling ages of WD components, plus an average estimate for MS lifetime, to explore the activity vs. age relation among lower MS stars.  This study confirmed that stars later than spectral type dM3 do not exhibit a Skumanich-style CA vs. age relation.  In a study of activity among unresolved WD+dM cataclysmic variable candidates Silvestri et al (2006) again used the white dwarf (WD) cooling times alone as lower age limit.  Both of these studies found the general trend seen in clusters, i.e, that later dM spectral types remain active at a roughly constant level for a longer period of time than earlier spectral types, whereupon each star becomes inactive.  The transition from active to inactive appears to take place over a relatively short period of time.  However, in both studies some dM stars in binaries were found to exhibit activity more characteristic of brighter, bluer and more massive M dwarfs than seen in clusters.  West et al. (2008) examined the age-activity relation among a sample of over 38,000 low-mass (M0-M7) stars drawn from the Sloan Digital Sky Survey (SDSS) Data Release 5 (DR5) and also found later spectral types remain active longer.  It is important to note that late type stars do not seem to decline in activity monotonically, they are either `on' if young, or `off' if old.  While the activity turnoff point on the lower MS constrains a cluster's age it is not a useful means for estimating the age of a field star.

This paper describes our analysis of a sample of common proper motion
binary (CPMB) systems. Most of the systems selected have a late MS star paired with a WD component. All have relatively wide orbital separations ($<a>$ $\sim$ 10$^{3}$ AU; Oswalt et al. 1993).
Thus, each component {is assumed to have} evolved independently, unaffected by mass exchange
or tidal coupling that complicates the evolution of closer pairs (Greenstein 1986). It is also assumed that members of such binaries are coeval. Essentially, each may be regarded as an open cluster with only two components. Such binaries are
far more numerous than clusters and span a much broader and more continuous range in age.
Thus, they are potentially very valuable to investigations of phenomena that depend upon age.
%for example, the age of a system can be estimated from the cooling age of the companion WD when that age far exceeds the main sequence lifetime of its progenitor. The WD ages are accurate to roughly 25\% (or better, if the mass
%of the WD is reliable), comparable to the accuracy determined
%from the main sequence turnoff ages of cluster.
The total age of a CPMB can be estimated from the cooling age of the WD component added to an estimate of its
progenitor's MS lifetime. Since some CPMBs in our sample
have ages well beyond the present $\sim$4 Gyr nearby cluster limit, this {provides an opportunity to test} the CA vs. age relation in much older MS stars.
%{\bf e.g. Silvestri et al. (2005) investigate the relationship between age and chromospheric activity for 139 M dwarf stars in CPMB system and find that
%M dwarfs in wide binaries older than 4 Gyr depart from the loglinear relation for clusters. }

Wide binaries present another opportunity. From the spectra of the MS components, the metallicity can be measured. Presumably this is also the original metallicity of the WD progenitor.
In addition, the radial velocities of the MS stars are easy to determine. Since proper motions are available, with trigonometric or photometric parallaxes, full space motions for all systems in the observed sample can be estimated. Thus, CPMBs present an opportunity to investigate the relations among age, metallicity and space motion, as well as population membership even for WDs, which often have weak or no spectral feature.

In section 2 we present an overview of the observations and
reductions for our sample. A discussion of the CA vs. age relation
is given in section 3. Our age and metallicity relation is presented
in section 4. In section 5, we describe our analysis of population membership. We conclude with a discussion of the implications of our findings in section 6.
% In this paper we present another approach based on cluster member stars and field stars which have accurate age data. % Most of our field stars are components of wide binaries which have a distant dwarf companion.
\section{Observations and Data Reduction}

%% In a manner similar to \objectname authors can provide links to dataset
%% hosted at participating data centers via the \dataset{} command.  The
%% second curly bracket argument is printed in the text while the first
%% parentheses argument serves as the valid data set identifier.  Large
%% lists of data set are best provided in a table (see Table 3 for an example).
%% Valid data set identifiers should be obtained from the data center that
%% is currently hosting the data.
%%
%% Note that AASTeX interprets everything between the curly braces in the
%% macro as regular text, so any special characters, e.g. "#" or "_," must be
%% preceded by a backslash. Otherwise, you will get a LaTeX error when you
%% compile your manuscript.  Special characters do not
%% need to be escaped in the optional, square-bracket argument.
 Most stars chosen for this study are components of wide MS+WD pairs  from the Luyten (1979) and Giclas, Burnham $\&$ Thomas (1971) proper motion catalogs chosen by Oswalt, Hintzen $\&$ Luyten (1988). A key impetus for using such pairs in this study is that the total lifetime of each pair should be approximately the age derived from measurements of
 the MS component. In addition the total age of a pair should be approximately the sum of the WD component's cooling time and the MS lifetime of its progenitor.

 Table 1 gives the observed data for 36 wide
binaries. Column 1 is a unique ID number. Columns 2 - 4 list each component's name, right ascension, and declination (coordinates are
for epoch 1950). The V magnitudes and original low-resolution
spectroscopic identifications are given in columns 5 - 8. Column 9 is the observation date for the high resolution ($\sim$ 2$\rm \AA$) spectra in the present study.
Columns 10 - 13 list the proper motion, direction of proper
motion (measured east of north), position angle (centered on the
primary measured east of north), and separation of the components (in arcseconds),
respectively. Of the 36 wide binaries, 5 systems consist of two MS stars, 1 system is a triple WD+dK+dM, and the remaining 30 are MS+WD pairs.

%The pairs chosen for this sample contain a WD within the color range 0 $\leq$ V-I $\leq$ 2.4 with a MS companion of spectral type earlier than dM3.  The latter limit was based on the findings of Silvestri et al. (2003; 2005) who found no Skumanich-style activity-age relation among stars later than dM3, in accord with the expectation that in the coolest MS stars a different so-called `turbulent dynamo' drives CA (See Reid et al. 2000 and reference therein).

A sample of cluster MS stars of known age, such as IC 2391, IC 2602 and M67, previously studied by
Patten $\&$ Simon (1993, 1996), Barrado y Navascu\'{e}s, Stauffer $\&$ Jayawardhana (2004) and Giampapa et al. (2006),
were adopted as `CA standards'.
These stars were routinely observed in the course of our observing program for the CPMBs.
Table 2 provides the observational information for these cluster member stars. Column 1 is a unique ID. Columns 2 - 4 list the name, V magnitude and observation date. The colors V-I and B-V are given in column 5 - 6.  The CA flux ratio S$\rm_{HK}$ (see Section 3.1), age and cluster membership are given in columns 7 - 9. The last column provides the corresponding literature source for each star.

\subsection{BVRI Photometry}
 We used BVRI photometric data for our wide binaries from Smith (1997) whenever
 available. Photometric data for some stars that were not in this source were taken from the literature identified
 by the Simbad Astronomical Database (Genova 2006). Photometric colors of other stars were estimated from our spectra by empirical calibrations. For example, Fig.1 is the relation between V-I and CaI4227 flux ratio. The feature of this index is within 4211 - 4242 $\rm\AA$ and the continuum ranges within 4152 - 4182 $\rm\AA$. The filled circles are stars with known V-I color. The dotted line is a least-squares fit. The scatter $\rm \sigma_{V-I} \approx $ 0.22. We used this relation to estimate the V-I for stars with no published colors. For the stars in cluster IC 2602, IC 2391 and M67, photometric
 colors were taken from Barnes, Sofia $\&$ Prosser (1999 and references therein), Patten $\&$ Simon (1996) and Giampapa et al. (2006).
\subsection{Spectroscopic Observations}
In the southern hemisphere, observations were conducted at Cerro Tololo Interamerican Observatory (CTIO) using
 the Blanco 4-meter telescope.  The Ritchey-Chr$\acute{e}$ti{e}n (RC) Cassegrain spectrograph was used on two
 separate observing runs (February 2004 and February 2005) to obtain optical spectroscopy of CPMBs, as well as the CA standard stars.
 During the two observation runs, the KPGL1 grating was used to obtain spectra with a scale of
 0.95 $\rm{\AA}$/pixel.  A Loral 3K CCD (L3K) was used with the RC spectrograph.
 It is a thinned 3K$\times$1K CCD with 15 $\mu$m pixels.  A spectral range of approximately 3800 $\rm{\AA}$
 to 6700 $\rm{\AA}$ was achieved.

 Northern hemisphere observations were conducted at Kitt Peak National Observatory (KPNO) using the Mayall 4-meter telescope. The RC spectrograph, with the BL450 grating set for the 2nd order to yield a resolution of 0.70 $\rm{\AA}$/pixel, was used to obtain optical spectra during the July 2005 and November 2006 observing runs. The 2K$\times$2K T2KB CCD camera with 24 $\mu$m pixels was used to image the spectra.  An 8-mm CuSO$_{4}$ order-blocking filter was added to decrease 1st-order overlap at the blue end of the spectrum. A spectral range of approximately 3800 $\rm{\AA}$ to 5100 $\rm{\AA}$ was achieved.
\subsection{Data Reduction}
The data were reduced with standard IRAF\footnote[1]{IRAF is distributed by the National Optical Astronomy
observatories, which are operated by the Association of Universitites for Research
in Astronomy, Inc., under cooperative agreement with the National
Science Foundation (http://iraf.noao.edu).} reduction procedures.
In all cases, program objects were reduced with calibration
data (bias, flat, arc, flux standard) taken on the same night. Data
were bias-subtracted and flat-fielded, and one-dimensional spectra
were extracted using the standard aperture extraction method. A
wavelength scale was determined for each spectrum using HeNeAr arc lamp
calibrations. Flux standard stars were used to place the spectra
on a calibrated flux scale. We emphasize that these are only relative fluxes, as most nights were
not spectrophotometric.

The radial velocity of each MS star was determined by cross-correlation between the observed spectra
and a set of MS template spectra. The F, G and K template spectra were generated from a theoretical atmosphere grid (Castelli $\&$ Kurucz 2003). The dM template spectra were compiled using observed
M dwarf spectra from the Sloan Digital Sky Survey (SDSS)\footnote[2]{http://www.astro.washington.edu/slh/templates}. The wavelength ranged from roughly 3900 - 9200 $\rm\AA$ (see Bochanski et al. 2007). Our typical internal measurement uncertainties in radial velocity
were $\sigma\rm_{v_{r}}$ = $\pm$ 4.6 km s$^{-1}$. The final radial velocities listed in column 5 of Table 3 were corrected to the heliocentric frame.

%% In this section, we use  the \subsection command to set off
%% a subsection.  \footnote is used to insert a footnote to the text.

%% Observe the use of the LaTeX \label
%% command after the \subsection to give a symbolic KEY to the
%% subsection for cross-referencing in a \ref command.
%% You can use LaTeX's \ref and \label commands to keep track of
%% cross-references to sections, equations, tables, and figures.
%% That way, if you change the order of any elements, LaTeX will
%% automatically renumber them.

%% This section also includes several of the displayed math environments
%% mentioned in the Author Guide.

\section{CA-Age Relation}
\subsection{Measurement of \rm{S$\rm_{HK}$}}
The flux ratio
\begin{eqnarray}
\rm{S_{HK}}& = & \alpha\rm{\frac{H+K}{R+V}}
\end{eqnarray}
was determined for each MS star, where H and K are the fluxes measured in 2 $\rm{\AA}$ rectangular
 windows centered on the line cores of CaII H$\&$K. Here R and V are the fluxes measured in 20 $\rm{\AA}$
 rectangular `pseudocontinuum' windows on either side. Although these are not strictly equivalent to
 the triangular windows Wilson (1968) used with his photomultiplier-based spectrometer, Hall, Lockwood $\&$ Skiff (2007) have shown that using 1.0 $\rm{\AA}$ rectangular H$\&$K windows produces results that are easily
 calibrated to the Baliunas et al. (1995) analysis of Wilson's (1968) original survey of bright MS stars.
 Our choice of 2 $\rm{\AA}$ windows is set by the resolution of the CTIO and KPNO instrumentation, but
 it detects CaII H$\&$K emission nearly as well and allows fainter stars to be observed. Gray et al. (2003) have shown that even a resolution of $\sim$4$\rm \AA$ can produce useful measure of S$\rm_{HK}$. In our measurement,
 the scale $\alpha$ is 10.0, reflecting the fact that the continuum windows are 10 times wider than the H$\&$K windows.

 In our sample some stars were observed two or three times. In such cases, the mean of these measurements was
 adopted as the star's $\rm{S_{HK}}$ and the scatter as the uncertainty. For those stars observed only once, we
 adopted the average uncertainty derived  from those stars having more than one observation ($\pm$4.6\%). The $\rm{S_{HK}}$ indices of all
 the MS stars are shown in column 4 of Table 3. For the purpose of this study we need only a calibration of CA vs. age on our instrumental system. Some stars were observed both at CTIO and KPNO. We found an empirical calibration: S$\rm_{ctio}$=S$\rm_{kpno}$+0.095. To remove this small instrumental effect, all the S$\rm_{HK}$ measured at KPNO were transformed into the CTIO instrumental system with this relation.
 %A comparison of
% our CA measurement to other systems will be performed in a future study as additional observations are collected.
\subsection{The age determination}
Our sample includes 14 cluster member stars: 6 in IC2602, 6 in IC2391 and 2 in M67. The ages of IC2602 and IC 2391
are approximately the same ($\sim$ 50$\pm$5 Myr; Barrado y Navascues, Stauffer $\&$ Jayawardhana 2004). They obtained intermediate-resolution optical spectroscopy of 44 potentially very low mass members  of IC2391 and derived the cluster age from a comparison of several theoretical models. The most recent age determination for M67 is 4.05$\pm$0.05 Gyr (Jorgensen $\&$ Lindegren 2005).
%In Giampapa et al. 2006, the cluster
%age was estimated from the mean chromospheric activity of its members.

  Among our 31 systems containing WD components, the ages of 23 were determined by using computed cooling times of WD companions added to estimates of their progenitor's MS lifetimes. The $T$$\rm{_{eff}}$ and log $g$ of each WD companion was obtained from the literature (see Table 4). The ages of the remaining 8 wide binaries were not obtained because the $T$$\rm{_{eff}}$ and log $g$ of WD companions could not be obtained from our spectra (3: DC type; 1: DQ; 4: low S/N) and could not be found in the literature. In cases where uncertainties were not given, we adopted 200 K and 0.05 as the average uncertainty for $T$$\rm{_{eff}}$ and log $g$, respectively. This decision was based on the recommendation of Bergeron, Saffer $\&$ Liebert (1992) who believe the internal errors are typically 100-300 K in $T$$\rm{_{eff}}$ and 0.02-0.06 in log $g$. From the $T$$\rm{_{eff}}$ and log $g$ of each WD, its mass (M$\rm_{WD}$) and cooling time (t$\rm_{cool}$) were estimated from Bergeron's cooling sequences\footnote[3]{The cooling sequences can be downloaded from the website: http://www.astro.umontreal.ca/~bergeron/CoolingModels/.}. For the pure hydrogen model atmospheres above $T$$\rm{_{eff}}$ = 30,000 K we used the carbon-core cooling models of Wood (1995), with thick hydrogen layers of q$\rm_{H}$ = M$\rm_{H}$/M$_{*}$ = 10$^{-4}$. For $T$$\rm{_{eff}}$  below 30,000 K we used cooling models similar to those described in Fontaine, Brassard $\&$ Bergeron (2001) but with carbon-oxygen cores and q$\rm_{H}$ = 10$^{-4}$ (see Bergeron, Leggett $\&$ Ruiz 2001). For the pure helium model atmospheres we used similar models but with q$\rm_{H}$ = 10$^{-10}$. M$\rm_{WD}$ and t$\rm_{cool}$ were then calculated by spline interpolation based on
the $T$$\rm{_{eff}}$ and log $g$.  In Table 4, columns 6 and 7 list the final M$\rm_{WD}$ and t$\rm_{cool}$ for these  23 WDs. Although $T$$\rm{_{eff}}$ and log $g$ are from different literature sources, the parameters are consistent for common stars in these references. Thus, there appears to be no systematic uncertainties expected among the $T$$\rm{_{eff}}$ and log $g$ we used.

Using the Initial-Final Mass Relation (IFMR; equations 2 - 3 presented in Catal\'{a}n et al. 2008b), we then estimated the progenitor masses M$\rm_{i}$ of the WDs.  There are two WDs whose masses are lower than 0.5M$_{\odot}$ and the current IFMR does not extend to such low mass objects. From Fig. 2 in Catal\'{a}n et al. (2008b), we adopted M$\rm_{i}$ $\sim$ 1.25M$_{\odot}$ for M$\rm_{WD}$ $<$ 0.5M$_{\odot}$. Next, using the third-order polynomial of Iben $\&$
Laughlin (1989),

\begin{eqnarray}
\rm{log~  t_{evol}}& =\rm{ 9.921 - 3.6648 (log M_{i}) + 1.9697 (log M_{i})^2 - 0.9369 (log M_{i})^3}
\end{eqnarray}

we determined the MS lifetime t$\rm_{evol}$ (in years) corresponding to each M$\rm_{i}$, progenitor mass of the WD (in
M$_{\odot}$). Columns 8 - 9 in Table 4 are the M$\rm_{i}$ and t$\rm_{evol}$ we derived for 23 WDs.
Finally, the total ages of these wide binaries were computed by adding t$\rm_{evol}$ to
the cooling ages of the WDs (t$\rm_{cool}$; column 11 in Table 3).

  Independent age determination for 4 pairs were found in the literature. These pairs are listed in Table 5. Column 1 gives their identifications. Columns 2-3 list the ages included in Holmberg, Nordstr\"{o}m $\&$ Andersen (2009) and Valenti $\&$ Fischer (2005), respectively. It can be seen that our ages are younger than isochrone fitting ages in all four cases. For 40 Eri A the isochone fitting age is unreasonably large, while our age of this star is consistent with the rotation age 4.75$\pm$0.75 Gyr from Barnes (2007). For CD-37 10500, the error bar is too large for isochrone fitting age to be useful. It is very difficult to estimate the age for K and M dwarfs with the isochrone fitting method because of the narrow vertical dispersion of isochrones within the MS in an HR diagram. Small uncertainties in luminosity and metallicity cause large uncertainties in age.   Using the white dwarf cooling times plus the estimated progenitor lifetimes should give a more consistent age for such stars. Note that the internal uncertainties of our age determinations are smaller than those derived from isochrone fitting.

   Although we have taken a similar approach, there are a few differences between our age determinations and those by Silvestri, Hawley $\&$ Oswalt (2005) and Silvestri et al. (2006). Silvestri et al. (2006) only used the WD cooling time as a lower limit to the age of each binary without considering the WD progenitor's lifetime in evaluating the age-activity relation among close binaries. Therefore, the ages were somewhat less than actual ages. In Silvestri,  Hawley $\&$ Oswalt (2005), each age was estimated from the WD's cooling time plus an estimate of its progenitor's lifetime. The M$\rm_{i}$ of each WD was computed from the IFMR of Weidemann(2000) using an adopted mean WD mass of 0.61 M$_{\odot}$. We explicitly computed the M$\rm_{i}$ of each WD from the IFMR of Catal$\rm \acute{a}$n (2008b) using our own estimate of its current mass (M$\rm_{f}$). In view of this, our method should provide more precise age estimates.
\subsection{Discussion}
All MS stars are plotted in Fig. 2, which displays our empirical relation among log(S$\rm_{HK}$), age and V-I. Asterisks represent the young cluster stars in IC2602 and IC2391 that have the same age 50$\pm$5 Myr (Barrado y Navascues, Stauffer $\&$ Jayawardhana 2004). It is clear that log(S$\rm_{HK})$ tends to be larger in red stars and smaller in blue stars at the same age. Two M67 cluster member stars are plotted with squares. M67 is much older than IC2602 and IC2391 and these points clearly demonstrate the expectation that CA declines with age. The plus ( + ) signs represent  field stars whose ages are between 1.0 Gyr and 5.0 Gyr. Diamonds represent stars whose ages are between 5.0 Gyr and 8.0 Gyr. Our sample includes five known flare stars (40 Eri C, LP891-13, LP888-63, G216-B14A, BD+26 730) which are displayed as filled circles. These clearly show enhanced CA for their supposed age.

Fig. 2 can be divided into four arbitrary age domains, represented by dotted lines. The top domain mainly consists of young active stars and flare stars. The other three domains consist of less active stars whose ages are, respectively,
%From top to bottom, we divide the log(S$\rm_{HK}$) domain into four parts by three dot lines. The first parts mainly consist of the young stars. The stars in this part are CA active stars. The other three parts are all CA inactive stars whose ages is older than 1 Gyr.
1.0 - 5.0 Gyr, 5.0 - 8.0 Gyr and $>$ 8.0 Gyr. The typical uncertainty in V-I and log(S$\rm_{HK}$) for inactive stars are displayed at the right-top of Fig. 2 as an error bar. A least-squares fit to stars in IC2602 and IC2391 (dash-dot line) illustrates that this group conforms to a consistent age in the log(S$\rm_{HK})$ vs. V-I plane.

Stars to the right of the vertical dashed line at V-I = 2.4 are later in spectral type than $\sim$dM3. Silvestri, Hawley $\&$ Oswalt (2005) found no Skumanich-style CA vs. age relation among stars later than dM3, in accord with the expectation that  a
so-called ¡®turbulent dynamo¡¯ drives CA in such stars (See Reid $\&$ Hawley 2000 and reference therein). Thus, stars in this region are not expected to follow the CA vs. age relation seen in earlier spectral type MS stars.

Overall, it can be seen that CA generally decays with age from 50 Myr to at least 8 Gyr for stars with 1.0 $\leq$ V-I $\leq$ 2.4. However, for stars bluer than V-I $\approx$ 1.0, CA shows little variation between 1.0-5.0 Gyr, which is consistent with the results of Pace et al. (2009) who found that CA from 1.4 Gyr (NGC3680) to 4.0 Gyr (M67) remains almost constant.  Also note that lines of constant age do not have the same slope. For young stars the lines of constant age have a relatively steep slope, while they appear to be much flatter for old stars.
%However, they also suggest that CA  decays almost stops after about 1.5 Gyr. If true, this would severely limit the use of CA as an age indicator. However, in Fig.1, even considering the uncertainty of log(S$\rm_{HK}$), it is clear that they continue to be a spread in decay
%after 1.5 Gyr.

Some comments on a few individual stars in Fig. 2 are appropriate.
 The star labeled as  1 in Fig. 2 is G163-B9A. Its companion, G163-B9B, was identified as an sdB star by Wegner
 $\&$ Reid (1991) and Catal\'{a}n et al. (2008a). The CA of G163-B9A is very strong. Thus, it is either a very young star or perhaps was observed during a flare event. The latter possibility was eliminated based on a detailed examination of its spectrum. Thus we conclude G163-B9A/B is probably not a physical pair.  Stars labeled as 2-3 are possible halo stars; star 4 has the weakest CA. These will be discussed  in Section 5.

There are 5 wide binaries (CD-31 1454/LP888-25; G114-B8A/B; CD-31 7352/LP902-30; LP684-1/2 and LP387-1/2) that consist of pairs of  MS stars. Only LP387-1/2 was observed at KPNO while all others were observed at CTIO. Since the two components of coeval binaries are expected to have consistent levels of CA, this provides a good reality check on our CA, V-I and age relation. Unfortunately, the spectrum of LP888-25, the companion to CD-31 1454, was unusable because of low signal-to-noise (S/N). In addition, we concluded that G114-B8A/B is not a physical binary because the components' radial velocities are inconsistent. The solid lines in Fig. 2 connect the MS components in the other three pairs CD-31 7352/LP902-30, LP684-1/2 and LP387-1/2, that most likely are physical pairs. The components in CD-31 7352/LP902-30 and LP387-1/2 have CA consistent with the same age. Some age difference is implied in LP684-1/2 though the difference is within the uncertainty implied by the error bar in the upper right corner of Fig. 2. Clearly, CA, along with radial velocity, provides a very useful way for filtering out nonphysical pairs.
%So it is feasible to get a rough age estimation from our CA, V-I and age relation.

\section{The age-[Fe/H] Relation}
The metallicity of each MS star was estimated by comparing the observed spectra to a set of template spectra. Initially, a library of low resolution theoretical spectra was generated using the SYNTH program, based on Kurucz's New ODF atmospheric models (Castelli $\&$ Kurucz 2003). The atmosphere models assume local thermodynamic equilibrium (LTE).
A mixing-length of l$/$H$\rm_{p}$ = 1.25 and a microturbulence $\xi$ = 2 km s$^{-1}$ were adopted. The line list included the atoms and molecules from Kurucz (1993). The maximum correlation method was applied to find the closest matching theoretical spectra to each observed spectrum. Since our objects are MS stars, we adopted log $g$ = 4.5 and [Fe/H] = 0.0 as initial values. The effective $T$$\rm{_{eff}}$ was then estimated based on the maximum correlation method. Once an estimated $T$$\rm{_{eff}}$ was obtained, a new estimated [Fe/H] could be obtained with the same procedure. After several iterations, the best-fitting parameters stabilized and computations were terminated. The left panel of Fig. 3 displays the correlation coefficient vs. [Fe/H] in a typical fit. Each open circle represents one template.
The filled circle is the maximum correlation coefficient obtained from polynomial fitting (solid line). The value of [Fe/H] at this point is regarded as our best estimate for this star's metallicity. The diamond points mark the templates yielding the maximum correlation (`+' : highest template point above; `-': the highest template point below). The difference in [Fe/H] between the two diamond points and the filled circle is taken as the estimated uncertainty in metallicity. The right panel of Fig. 3 compares our [Fe/H] estimates to those which could be found in the literature. The mean difference is smaller than  0.15 dex, suggesting our metallicity is basically consistent with that of other work. [Fe/H] measurements were not made for stars later than M3 because we do not have templates for the latest spectra nor could we expect the CA vs. age determination to be valid in such late type MS stars. The resulting metallicities estimated for 37 MS stars in our sample are given in column 10 of Table 3.

Although one of the key consequences of the stellar evolution
theory is the gradual increase in the metal content of the
interstellar medium (ISM) and the progressive enrichment of
subsequent stellar generations, some studies have found little,
if any, indication that an age-metallicity relation (AMR) exists
amongst solar neighborhood late-type stars. For example, Rocha-Pinto et al. (2000) studied the AMR using a sample of 552 late-type dwarfs. For those stars, metallicities were estimated from \textit{uvby} data, and
ages were calculated from their chromospheric emission levels
using a metallicity-dependent CA vs. age
relation developed by Rocha-Pinto $\&$ Maciel (1998). The resulting AMR was found to be a smooth function in their analysis.  Feltzing, Holmberg $\&$ Hurley (2001) found that the solar neighborhood age-metallicity diagram is well populated at all ages in a sample of 5828 dwarf and sub-dwarf stars from the Hipparcos Catalogue. Bensby, Feltzing $\&$ Lundstr\"{o}m (2004) investigated the AMR using a sample of 229 nearby thick disk stars. Their results indicate that there is indeed an AMR in
the thick disk. They found that the median age
decreases by about 5-7 Gyr when going from [Fe/H]$\approx$ -0.8 to [Fe/H]$\approx$ -0.1.

Fig. 4 is our [Fe/H] vs. age relation for 21 stars. An asterisk represents one likely halo star (see next section). Disk stars are displayed as filled circles. The typical uncertainties in age and [Fe/H] are shown at the left-bottom of this figure. The dotted line is a least-squares fit for only disk stars, while the dashed line is the fit for all stars. The expected trend in age-metallicity is apparent: old stars tend to be of lower metallicity.

 Our [Fe/H] vs. age relation differs somewhat from the early work on single stars by Barry (1988; see Fig. 5 in that reference). Our Fig. 4 shows a more clear trend, quite similar in fact to the newer study of field stars by Rocha-Pinto $\&$ Maciel (2000); the slope of their relation and ours are approximately the same. We conclude that wide binaries exhibit a [Fe/H] $\sim$ age relation similar to field stars. However, at present our sample contains too few stars to support a detailed examination of the nearby star formation history.

%Fig. 4 shows a plot of $\delta$S vs. [Fe/H] for 43 of our stars (excluding stars later than dM2, some very active stars and stars with large [Fe/H] uncertainty). The dashed-dotted line is least squares fit. Overall maybe a very weak tendency for $\delta$S to decrease with metallicity. To better known the [Fe/H] effect, we divide Fig. 3 into four parts (see the dot line). Active stars are all in Part I, where there is nearly no relation (the Dash line). The metallicity scatter is large in this part. Metal-rich stars are in Part II.  Some stars show lower activity approximately resulted from the stronger Ca features. There is an clear relation (the Dash line) in part III. The metal poor stars have lower CA means that they are old, which is consistent with our understanding of the age-metallicity-relation (AMR). Only two metal-poor stars are in part IV, it is hard to illustrate the trend in this zone.

\section{Population Membership}
Column 6 of Table 3 lists the parallaxes of our wide binaries. Some trigonometric parallaxes were obtained from the Simbad Astronomical Database (Genova 2006). For 11 wide binaries lacking trigonometric parallaxes, we computed photometric parallaxes using equation 3-5 which were derived from the data in Bergeron, Leggett $\&$ Ruiz (2001).

\begin{eqnarray}
\rm{\pi}& =\rm{10^{-(\frac{V-M_{v}+5}{5}})}
\end{eqnarray}

\begin{eqnarray}
\rm{M_{v}}& =\rm{ 12.2199 + 1.8152(V-I) + 2.9704(V-I)^2 - 1.7082(V-I)^3}
\end{eqnarray}

\begin{eqnarray}
\rm{M_{v}}& =\rm{ 11.7099 + 6.6038(B-V) - 3.7273(B-V)^2 + 0.8066(B-V)^3}
\end{eqnarray}

The rectangular velocity components relative to the Sun
for 41 MS stars were then computed and transformed
into Galactic velocity components U, V, and W,
and corrected for the peculiar solar motion
(U, V, W) = (-9, +12, +7) km s$^{-1}$ (Wielen 1982). The UVW-velocity
components are defined as a right-handed system with U positive
in the direction radially outward from the Galactic center,
V positive in the direction of Galactic rotation, and W
positive perpendicular to the plane of the Galaxy in the
direction of the north Galactic pole. The typical uncertainties in U, V and W are no
more than $\sim$10 km s$^{-1}$. Columns 7-9 of Table 3 list our computed U, V and W velocity components.

The top panel of Fig. 5 shows contours, centered at
(U, V) = (0, -220) km s$^{-1}$, that represent 1$\sigma$ and 2$\sigma$ velocity ellipsoids
for stars in the Galactic stellar halo as defined by Chiba $\&$
Beers (2000).  Six stars lie outside the 2$\sigma$ velocity contour centered on (U, V) = (0, -35) km s$^{-1}$ defined for disk stars (Chiba $\&$ Beers 2000). The bottom of Fig. 5 shows the Toomre diagram of our stars. Venn et al. (2004) suggest stars with V$_{total}$ $>$ 180km s$^{-1}$ are possible halo members. There are two stars that meet this criterion. Taking metallicity, space motion and CA into account, they have a high probability of belonging
to the halo population.  One is LHS300A ([Fe/H]= -0.95 $\pm$ 0.25) which is identified as a thick disk star in Monteiro et al. (2006). Considering the metallicity and space motion, we think it is a halo star. The other is CD-31 1454 ([Fe/H] = -0.48 $\pm$ 0.02) which is regarded as a halo star by Chanam\'{e} $\&$ Gould (2004).  The two likely halo stars are also labeled in Fig. 2 as number 2 and 3 respectively. Their CA is weak, suggesting ages in excess of 5 Gyr. These two plausible halo stars are displayed as asterisks plus filled circles in the bottom pane of Fig. 5. The star numbered 4 in Fig. 2 is G114-B8B ([Fe/H] = -0.40 $\pm$ 0.08) which may be the oldest star in our sample. Its age appears to be older than 8 Gyr as suggested by its location in Fig. 2. The other stars are most likely members of the disk. The above analysis demonstrates the difficulty of making a population assignment on the basis of only space velocity or metallicity or CA. Ideally all three should be used.

The left panel of Fig. 6 displays the computed absolute value of the W components of our stars' space motions vs. estimated ages for 21 stars for which complete information is available. As expected, old stars tend to have larger W velocity. A weak positive correlation between the vertical velocity (W) of stars in CPMBs and age is expected based on the standard paradigm for stellar encounters in the disk. The right panel of Fig. 6 presents the \textit{dispersion} in W as a function of age. It is clear that in general the W dispersion becomes larger with age from 1 Gyr to 8 Gyr.

\section{Conclusion}
In this study we presented the CA levels, ages, metallicities and space motions for components of 36 wide binaries. WD components were identified in 31 wide binaries. We also observed a sample of cluster member stars with well-determined ages in order to test the expected  CA vs. age relation. The ages of 23 wide binaries were derived by the cooling time of each WD companion added to the lifetime estimate of its progenitor.

We first examined the relation among log (S$\rm_{HK}$), V-I and age. Our results support the expected
hypothesis established among single and cluster MS stars, i.e., in general CA declines with age for stars with 1.0 $\leq$ V-I $\leq$ 2.4 from 50 Myr to at least 8 Gyr. However, for stars with V-I $<$ 1.0, the CA varies little between 1 Gyr to 5 Gyr. This is consistent with results of Pace et al. (2009) who found nearly constant CA level from 1.4 Gyr (NGC3680) to 4.0 Gyr (M67). Apparently the slope in the log(S$\rm_{HK}$) vs. V-I plane for young stars is relatively steep, while for old stars it appears to be flatter. Additional observations will be required to determine whether this slope changes monotonically or discontinuously with age.
These limitations will need to be taken carefully into account by anyone attempting to use CA to determine ages for single stars.

The metallicities of stars earlier in spectral type than M3 were measured by template matching. Our sample generally supports the expected paradigm, i.e. older stars tend to have lower metallicity. However, it also underscores the fact that there is much variation in metallicity at all ages, precluding its use for determining ages for single stars. Also, the AMR among wide binaries appears to be quite comparable to that found in single field stars (Rocha-Pinto $\&$ Maciel 2000).
%In addition, the scatter in metallicity about the mean at a given age tends to increase with age.

With trigonometric parallaxes from the literature and photometric parallaxes derived from B, V, R, I data, proper motions and our measured v$_{r}$ values, we calculated full space motions U, V and W for as many of our MS stars as possible. Our results clearly show that the W dispersion increases with age. In general, the W velocity component is also relatively larger for old stars. Using our measurements of CA and metallicity, we concluded that 2 wide binaries ( CD-31 1454/LP888-25 and LHS300A/B) are probably halo members, while the others are disk stars. This low fraction of halo members among wide binaries is consistent with the earlier results of Silvestri, Oswalt $\&$ Hawley (2002) who found only one halo binary in their sample of WD+dM pairs. We estimated the oldest disk star in our sample is G114-B8B ($>$ 8 Gyr) based on its weak CA.

 Five of our wide binaries consist of two MS stars. Two (G114-B8A/B; G163-B9A/B) apparently are not physical pairs, since the two companions have inconsistent CA levels, radial velocities and/or metallicities. CD-31 7352/LP902-30 and LP387-1/2 are physical pairs because the two MS components have very similar velocities, metallicty and CA level. LP684-1/2 is probably a physical pair since the radial velocity difference between two components is within the range of uncertainty and its components display comparable S$\rm_{HK}$ value.

 In conclusion, our study affirms the assumption that wide binaries (CPMBs) share the same kinematic $\&$ spectroscopic properties as nearby single field and cluster stars. Thus they are very promising resources for studying stellar populations and age groups that are not well sampled by nearby clusters.

%
%The $\delta$S and  metallicity relation was examined. Even though the metal-rich stars tend to be less active, there may be several regimes in the [Fe/H] vs. $\delta$S plot that will need to be understood before assuming a monotonic CA vs. age relation for all types of stars. More data are needed to better understand how metallicity affects the CA-age relation. Our preliminary assumption is that there is NO correlation between $\delta$S and [Fe/H]. It's a scatter plot.

%% The displaymath environment will produce the same sort of equation as
%% the equation environment, except that the equation will not be numbered
%% by LaTeX.

%% If you wish to include an acknowledgments section in your paper,
%% separate it off from the body of the text using the \acknowledgments
%% command.

%% Included in this acknowledgments section are examples of the
%% AASTeX hypertext markup commands. Use \url without the optional [HREF]
%% argument when you want to print the url directly in the text. Otherwise,
%% use either \url or \anchor, with the HREF as the first argument and the
%% text to be printed in the second.

\acknowledgments
TDO acknowledges supported from NSF grant AST-0807919 to Florida Institute of Technology. JKZ, GZ and YQC  acknowledge  support from NSFC grant No. 10821061, 11073026 and 11078019. We are also grateful for constructive comments by the reviewer that substantially improved our paper.

\clearpage

%% Use the figure environment and \plotone or \plottwo to include
%% figures and captions in your electronic submission.
%% To embed the sample graphics in
%% the file, uncomment the \plotone, \plottwo, and
%% \includegraphics commands
%%
%% If you need a layout that cannot be achieved with \plotone or
%% \plottwo, you can invoke the graphicx package directly with the
%% \includegraphics command or use \plotfiddle. For more information,
%% please see the tutorial on "Using Electronic Art with AASTeX" in the
%% documentation section at the AASTeX Web site,
%% http://www.journals.uchicago.edu/AAS/AASTeX.
%%
%% The examples below also include sample markup for submission of
%% supplemental electronic materials. As always, be sure to check
%% the instructions to authors for the journal you are submitting to
%% for specific submissions guidelines as they vary from
%% journal to journal.

%% This example uses \plotone to include an EPS file scaled to
%% 80% of its natural size with \epsscale. Its caption
%% has been written to indicate that additional figure parts will be
%% available in the electronic journal.

%\begin{figure}
%\epsscale{.80}
%\plotone{f1.eps}
%\caption{Derived spectra for 3C138 \citep[see][]{heiles03}. Plots for all sources are available
%in the electronic edition of {\it The Astrophysical Journal}.\label{fig1}}
%\end{figure}

\clearpage

%% Here we use \plottwo to present two versions of the same figure,
%% one in black and white for print the other in RGB color
%% for online presentation. Note that the caption indicates
%% that a color version of the figure will be available online.
%%
\begin{figure}
\epsscale{.80}
\plotone{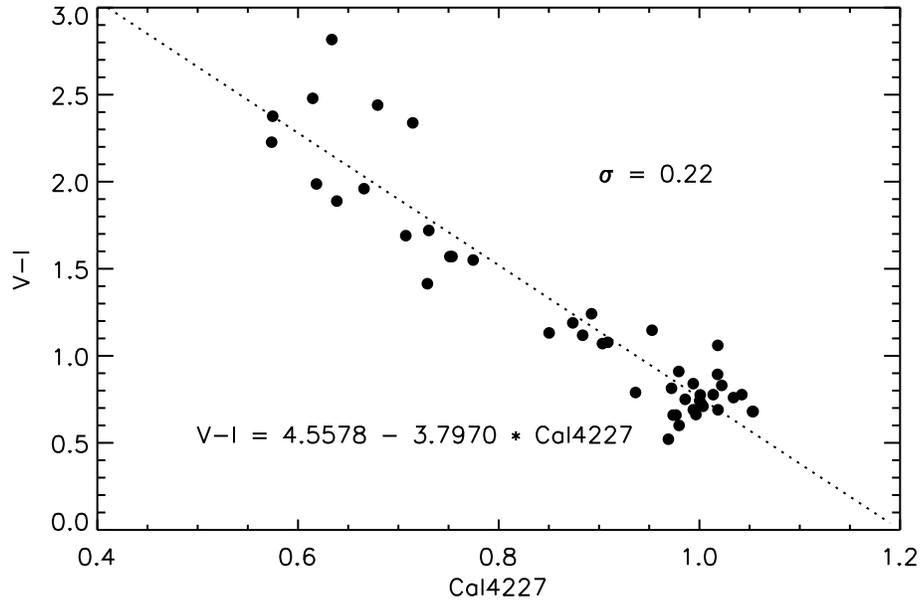}
\caption{The relation between V-I and flux ratio of CaI4227. The dotted line is a least-squares fit. We used this relation to estimate V-I color for a few stars whose photometric data was unavailable. See text for details.}
\end{figure}

\begin{figure}
\epsscale{.80}
\plotone{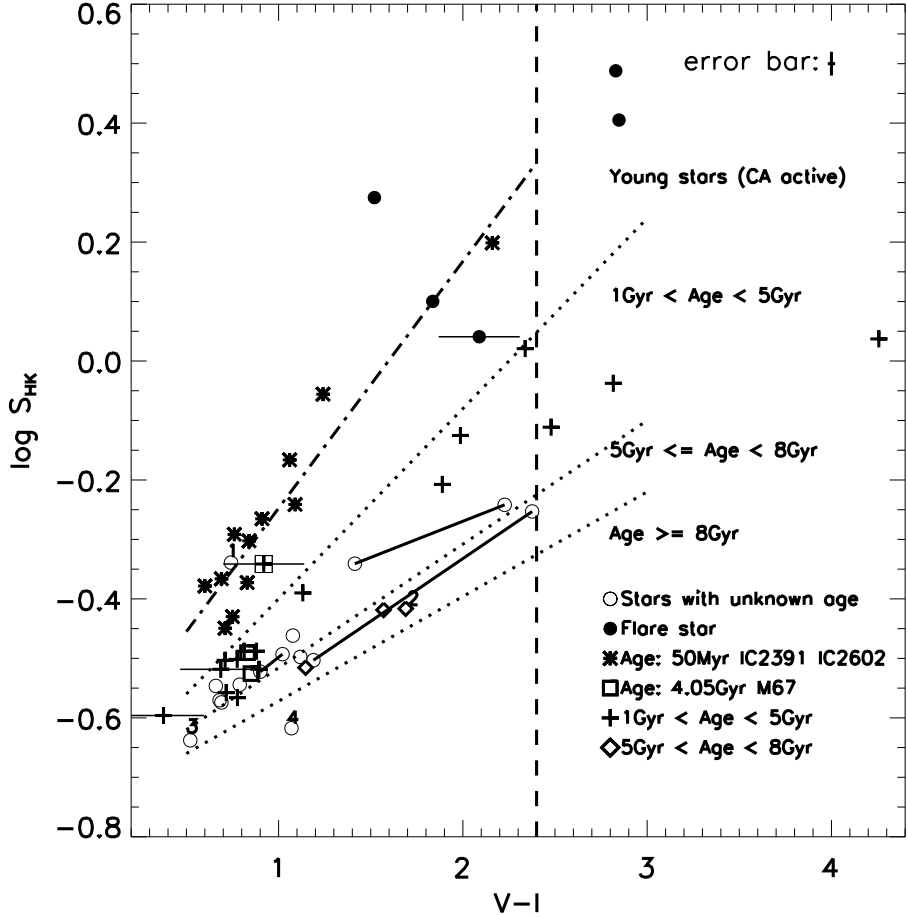}
\caption{The relation among log($\rm{S_{HK}}$),  V-I and age. Open circles represent CPMB stars with unknown age. Filled circles are flare stars. Stars in cluster IC2602 + IC2391 and M67 are shown as asterisks and squares respectively. Plus signs and diamonds represent stars with different ages: 1Gyr $<$ age $<$ 5Gyr and 5Gyr $<$ age $<$ 8Gyr, respectively. Four domains are divided by dotted lines. The dash-dot line is the least-squares fit of young clusters IC2602+IC2391. The vertical dashed line is V-I = 2.4. To the right of this, CA is not expected to depend on age. Solid lines connect the MS components in CD-31 7352/LP902-30, LP684-1/2 and LP387-1/2.  {The star indicated as number 1 is G163-B9A. Numbers 2-3 denote halo star candidates (LHS300A and CD-31 1454). Number 4 is the star with the weakest activity in our sample (G114-B8B)}. The typical error bar for our measurement is in the upper right corner. Separate V-I uncertainties are given for 4 stars whose V-I were estimated by the relation in Fig.1. A square with a plus sign is an outlier (G95-B5A) that seems overactive for its age domain. This star is perhaps a close binary and it warrants follow-up observations to determine the cause of its high CA.}
\end{figure}

\begin{figure}
\epsscale{1.2}
\plottwo{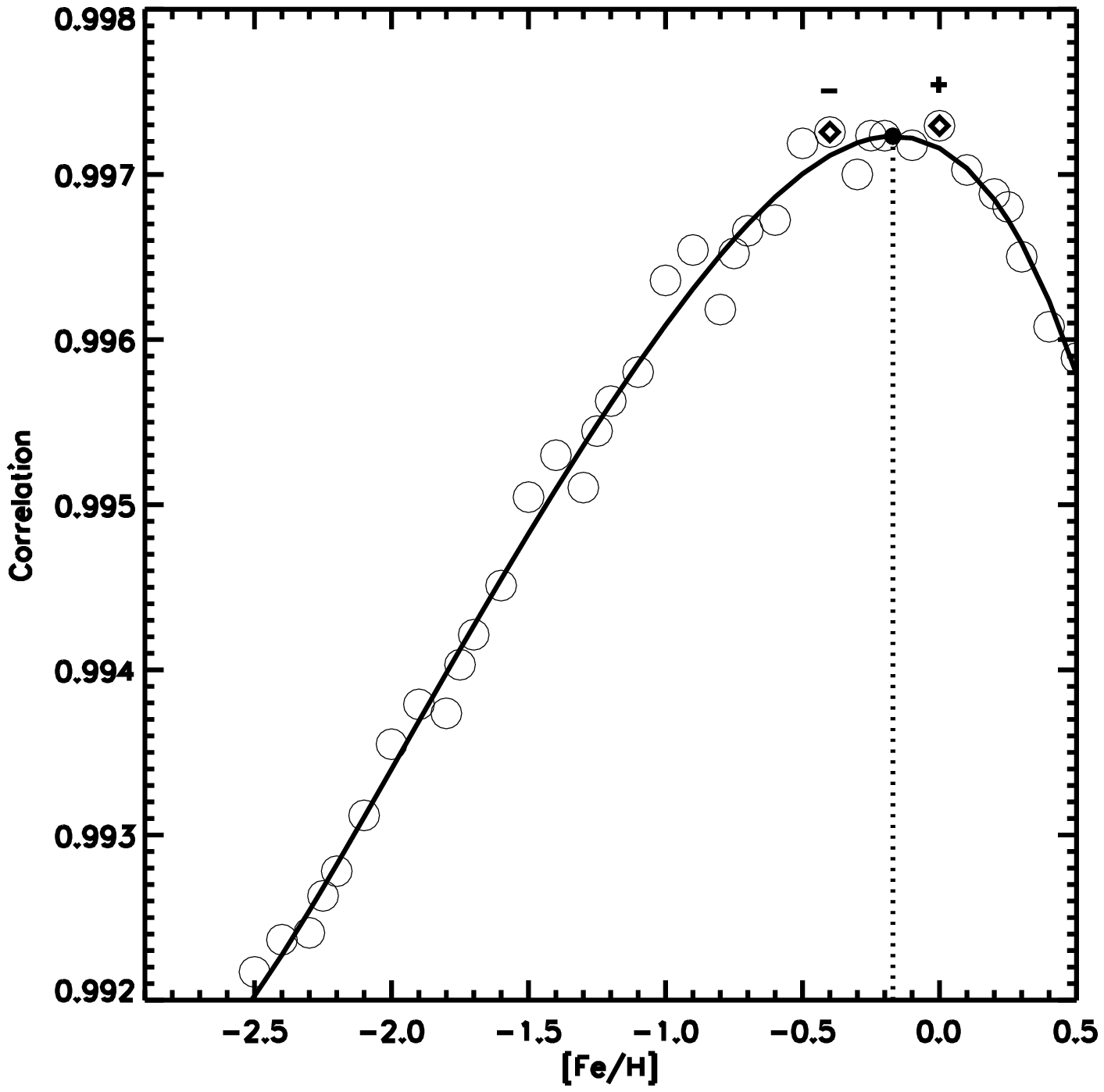}{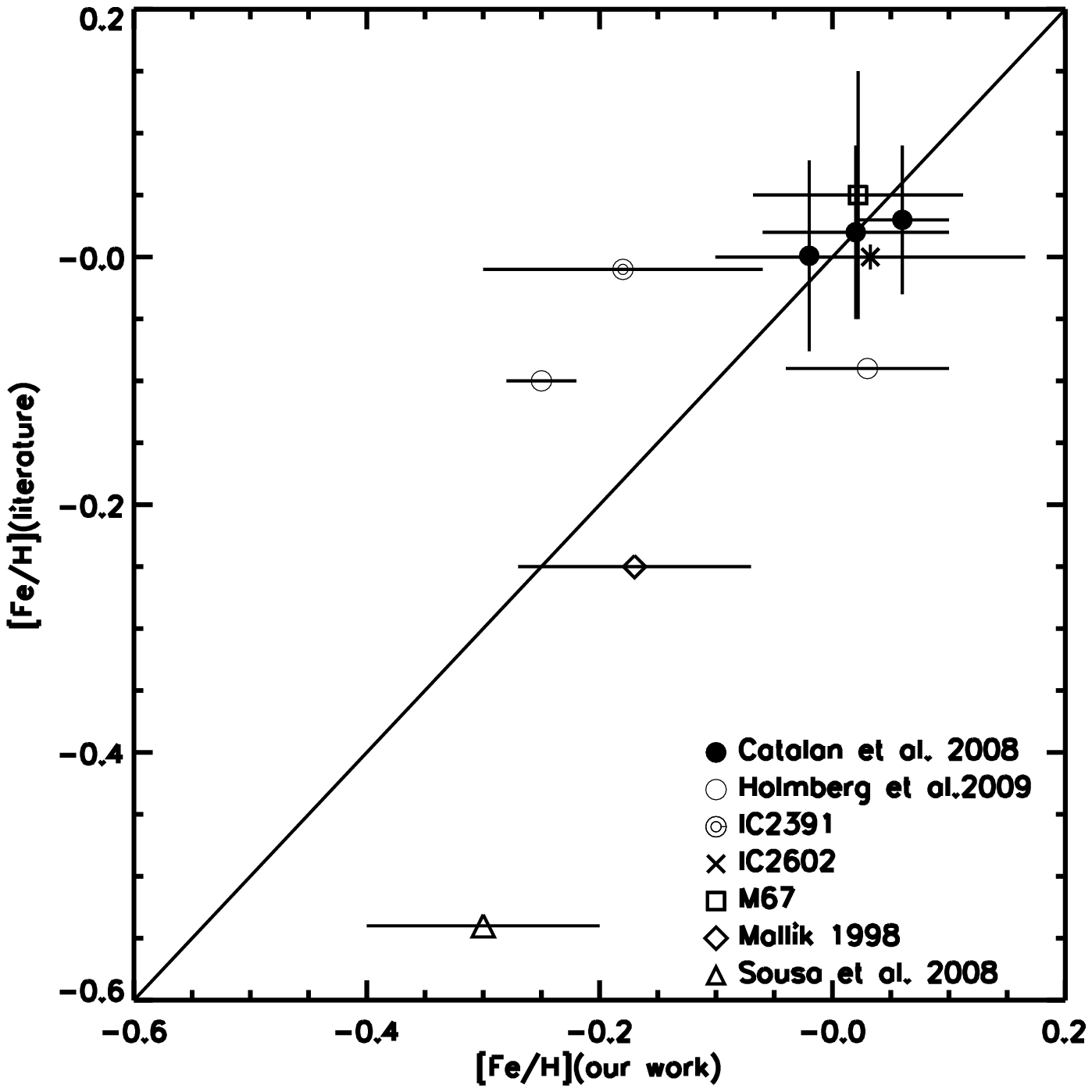}
\caption{Left: An example of how [Fe/H] estimates were made with the maximum correlation method. The filled circle is the maximum of fitted polynomial. The [Fe/H] correlated to this maximum is regarded as our final result. + and - signs indicate our estimate of uncertainty as discussed in the text. Right: The comparison between our [Fe/H] results and those in the literature. Filled circles represent stars in Catal\'{a}n et al. 2008a. Single open circles represent stars in Holmberg, Nordstr\"{o}m $\&$ Andersen (2009). Double open circles, X signs and squares represent the clusters IC2391, IC2602 and M67, respectively. We adopted the average [Fe/H] of member stars as the final cluster [Fe/H].  The [Fe/H] of IC2391 and IC2602 are from D'Orazi $\&$ Randich (2009), while that of M67 is from Pancino et al. 2010. Diamonds and triangles represent stars in Mallik (1998) and Sousa et al. (2008), respectively. Note that many of these sources did not give uncertainties, hence no vertical error bar could be plotted. }
\end{figure}

\begin{figure}
\epsscale{.8}
\plotone{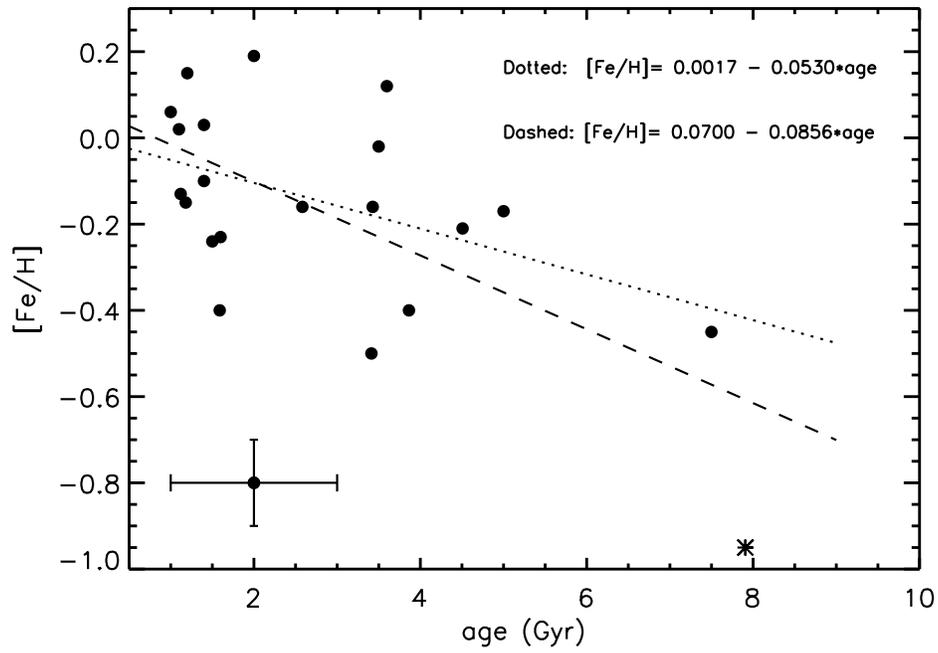}
\caption{Our [Fe/H] vs. age relation for 21 MS stars.  Typical uncertainties in age and [Fe/H] are displayed at the lower left of this figure. Filled circles represent disk stars. Only one star indicated by an asterisk (LHS300A) is a likely halo star. The dotted line is  a least-squares fit for only disk stars, while the dashed line includes all stars.}
\end{figure}

\begin{figure}
\epsscale{.8}
\plotone{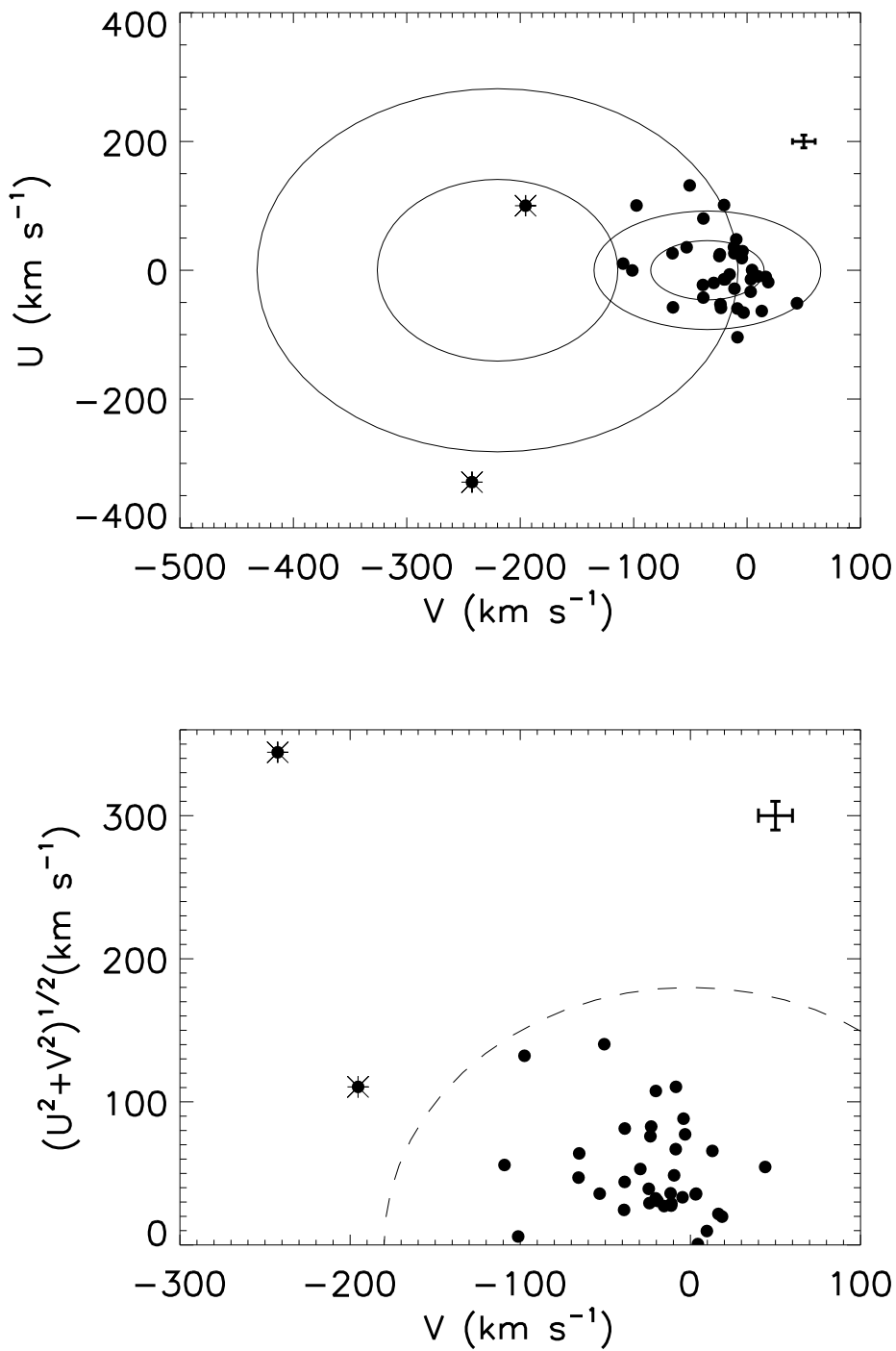}
\caption{Top: UV-velocity distribution of our sample with measured v$_{r}$ (km s$^{-1}$). The ellipsoids indicate the 1 $\sigma$ (inner) and 2 $\sigma$ (outer) contours
for Galactic thick-disk and stellar halo populations, respectively.  Typical error bars are given in both panels.
 Bottom: Toomre diagram of our stars. Dashed line is V$_{total}$ = 180 km s$^{-1}$. }
\end{figure}

\begin{figure}
\epsscale{.8}
\plotone{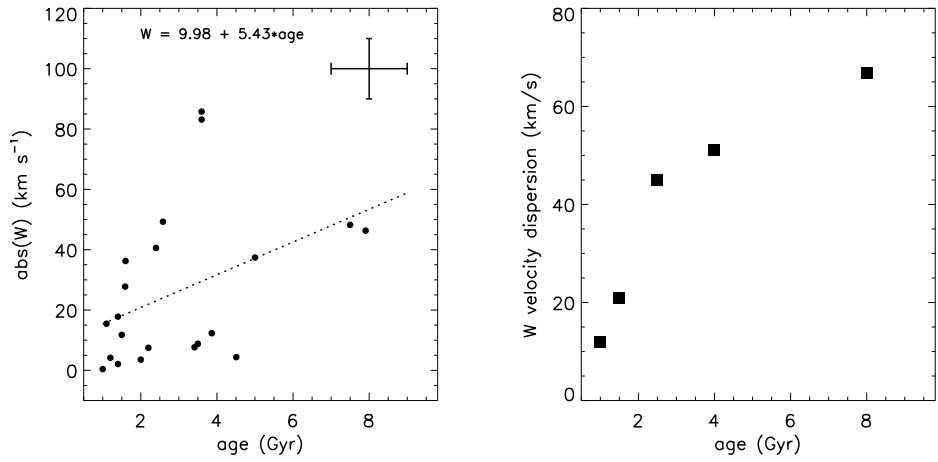}
\caption{Left: age vs. absolute value of the W component of space velocity for 21 MS stars in our CPMB sample. Solid line is a least-squares
fit. Typical uncertainties in age and W are displayed at the upper right of this figure. Right: W dispersion vs. age derived by binning the data in the left panel in five age group ranging from 1-8 Gyr. }
\end{figure}

\clearpage

\begin{deluxetable}{ccccccccccccccc}
\tabletypesize{\scriptsize}
\setlength{\tabcolsep}{0.02in}
\tablewidth{0pt}
%\rotate
\tablecaption{ Observed wide binaries  \label{tbl-1}}
\tablehead{
\colhead{ID}&\colhead{Name} & \colhead{R.A.}& \colhead{Decl.}& \colhead{} & \colhead{} &\colhead{} &\colhead{} &\colhead{UT} & \colhead{$\mu$}& \colhead{$\theta_{\mu}$}& \colhead{pos}& \colhead{Sep} & \colhead{Location}\\
& (Star 1/Star 2)&(B1950.0)&(B1950.0)&V1&Sp1\tablenotemark{a}&V2&Sp2\tablenotemark{a}&(mm/yy)&(arcsec yr$^{-1})$&(deg)&(deg)&(arcsec) \\
(1)&(2)&(3)&(4)&(5)&(6)&(7)&(8)&(9)&(10)&(11)&(12)&(13) \\
}
\startdata
 1&CD-31 1454/LP888-25&03 32 36& -31 14 00&11.8& dG&15.9&dK&02/05&0.51&186&264&223&CTIO \\
 2&G114-B8A/B         &08 58 17& -04 10 24&11.0&dK0&15.9&dG2&02/05&0.10&150&151&54&CTIO \\
 3&BD+12 937/G102-39  &05 51 05&  12 23 48& 7.6&dF8&15.7&DC &02/04&0.28&184&47 &91&CTIO \\
 4&G272-B5A/B         &02 00 31& -17 07 30&12.5&dG &15.8&DA &02/05&0.05&181&79&72 &CTIO \\
 5&BD-03 2935/LP670-9 &10 27 24& -03 57 00&11.2&dG &18.7&   &02/05&0.18&103&139&35 &CTIO\\
 6&G163-B9A/B         &10 43 39& -03 24 06&12.6&dF9&15.6&   &02/05&0.08&115&122&76 &CTIO\\
 7&BD+23 2539/LP378-537&13 04 48& 22 43 00&9.8 &dK0&16.2&DA &02/04&0.11&300&106&20 &CTIO\\
 8&CD-25 8487/LP849-59&11 07 00& -25 43 00&9.3 &sdM0&16.8&DC&02/04&0.25&106&181&100 &CTIO\\
 9&CD-28 3361/LP895-41&06 42 34& -28 30 48&11.2&dK  &16.8&DA&02/04&0.16&227&75&16 &CTIO  \\
10&BD-5 3450/LP674-29 &12 09 48& -06 05 00&12.0&dK5 &17.2&DC&02/05&0.44&220&102&202 &CTIO\\
11&BD-18 2482/LP786-6 &08 45 18& -18 48 00&12.8&dK3 &15.1&DB&02/04&0.16&268&236&31  &CTIO\\
12&40 Eri A/B/C        &04 13 03& -07 44 06&5.3 &dG  &9.5 &DA&02/04,02/05&4.08&213&105&82&CTIO \\
13&CD-31 7352/LP902-30&09 28 20& -31 53 12&9.3 &dK  &14.5&dM&02/05&0.34&348&206&12 &CTIO \\
14&LP684-1/2          &15 54 00& -04 41 00&12.7&dM&  15.5&dM&02/05&0.32&244&202&5   &CTIO\\
15&BD-18 3019/LP791-55&10 43 30& -18 50 00&12.9&dM0& 16.6&DQ&02/04&1.98&250&356&7.5 &CTIO\\
16&LP856-54/53        &13 48 30& -27 19 00&13.9&dM &15.1&DA&02/04&0.24&166&233&9    &CTIO\\
17&LP498-25/26        &13 36 45&  12 23 48&13.9&dM&14.5&DB&02/04&0.19&134&307&87 &CTIO\\
18&LP672-2/1          &11 05 30& -04 53 00&12.6&dM6&13.8&DA&02/04&0.44&184&160&279 &CTIO\\
19&LP916-26/27        &15 42 18& -27 30 00&15.5&dM&16.3&DB&02/04&0.24&235&330&52  &CTIO\\
20&LP891-13/12        &04 43 18& -27 32 00&15.6&dM&15.9&DQ&02/05&0.24&246&62&49   &CTIO\\
21&LP783-2/3          &07 38 02& -17 17 24&12.9&dM&17.6&DB&02/04,02/05&1.26&117&276&21 &CTIO \\
22&CD-37 10500/L481-60&15 44 12& -37 46 00&6.8&dG&13.2&DA&02/05&0.48&243&131&15   &CTIO\\
23&CD-59 1275/L182-61 &06 15 36& -59 11 24&7.0&dG0&13.7&DB&02/04,02/05&0.33&190&302&41 &CTIO\\
24&LP888-63/64        &03 26 45& -27 18 36&13.9& &15.6& &02/05& 0.83&63&227&7   &CTIO\\
25&CD-38 10983/10980  &16 20 38& -39 04 42&6.1&dG&10.7&DA&02/04,02/05&0.08&95&248&345&CTIO\\
26&LHS193A/B          &04 30 50& -39 08 55&11.7&dM&17.7 &DB&02/05& 1.023   &44.5  &    & &CTIO  \\
27&LHS300A/B          &11 08 58& -40 49 05&13.2    &dK& 17.8  &DB&02/05&1.277 &264.5  &    &   &CTIO\\
28&LP387-2/1          &16 44 18&  24 06 00&16.8&dG&17.6&DG&07/05&0.11&163&296&37 &KPNO \\
29&BD-8 0980/G156-64  &22 53 12& -08 05 24&9.0&dG&16.4&DA&07/05&0.59&91&168&43 &KPNO \\
30&G171-62/G172-4     &00 30 17&  44 27 18&10.3&dK&16.6&DA&11/06&0.16 &285.9&&&KPNO \\
31&BD-1 469/LP592-80 &03 15 48&	-01 06 18&6.6&dG&17.2&DA&11/06&0.18&192&50&49&KPNO\\
32&G216-B14A/B& 22 58 49&	40 40 12&12.0&&15.5&&11/06&0.07&215&261&23&KPNO\\
33&BD+44 1847/G116-16&09 11 51 &	44 15 36&10.2&&15.5&&11/06&0.28&174&95&1020&KPNO\\
34&G273-B1A/B&23 50 54&	-08 21 06&12.0&dG&16.4&DA&11/06&0.12&75&210&36&KPNO\\
35&G95-B5A/B& 02 20 44&	22 14 00& 9.2&&15.6&&11/06&0.15&113&94&26&KPNO\\
36&BD+26 730/LP358-525&04 33 42&	27 02 00&9.4&dK&16.3&DA&11/06&0.28&122&338&128&KPNO\\
\enddata
\tablecomments{Units of right ascension are hours, minutes, and seconds, and units of declination are degrees, arcminutes, and arcseconds.  $V$ magnitude values (except 26 and 27) are $m_{pg}$ magnitudes from Oswalt, Hintzen $\&$ Luyten 1988.}
\tablenotetext{a}{Spectral types for the WD and MS stars were determined from low - resolution ($\sim$ 7 - 15 $\rm{\AA}$) spectra (Oswalt et al. 1988, 1991, 1993).}

\end{deluxetable}

\begin{deluxetable}{cccccccccc}
\tabletypesize{\scriptsize}
%\rotate
\tablecaption{The list of cluster member stars\label{tbl-2}}
\tablewidth{0pt}
\tablehead{
\colhead{ID}&\colhead{Name} &\colhead{V}& \colhead{UT}& \colhead{V-I}& \colhead{B-V}& \colhead{S$\rm_{HK}$}&\colhead{age (Gyr)}&\colhead{Cluster} &\colhead{Ref\tablenotemark{a}} \\
(1)&(2)&(3)&(4)&(5)&(6)&(7)&(8)&(9)&(10) \\
  }
\startdata
 37&[RSP95] 15&11.75&02/05&1.0600&0.9300&0.6819&0.050$_{0.045}^{0.055}$&IC2602&1 \\
 38&[RSP95] 32&15.06&02/05&2.1600&1.6300&1.5799&0.050$_{0.045}^{0.055}$&IC2602&1 \\
 39&[RSP95] 66&11.07&02/04&0.8300&0.6800&0.4241&0.050$_{0.045}^{0.055}$&IC2602&1 \\
 40&[RSP95] 70&10.92&02/04&0.7100&0.6900&0.3553&0.050$_{0.045}^{0.055}$&IC2602&1 \\
 41&[RSP95] 72&10.89&02/05&0.7600&0.6400&0.5110&0.050$_{0.045}^{0.055}$&IC2602&1 \\
 42&[RSP95] 80&11.75&02/04&1.0900&0.9300&0.5738&0.050$_{0.045}^{0.055}$&IC2602&1 \\
 43&VXR PSPC 12&11.86&02/04&0.9100&0.8300&0.5429&0.050$_{0.045}^{0.055}$&IC2391&2 \\
 44&VXR PSPC 14&10.45&02/04&0.6900&0.5700&0.4302&0.050$_{0.045}^{0.055}$&IC2391&2 \\
 45&VXR PSPC 70&10.85&02/04&0.7500&0.6400&0.3716&0.050$_{0.045}^{0.055}$&IC2391&2 \\
 46&VXR PSPC 72&11.46&02/04&0.8400&0.7300&0.4984&0.050$_{0.045}^{0.055}$&IC2391&2 \\
 47&VXR PSPC 76a&12.76&02/04&1.2414&1.0400&0.8793&0.050$_{0.045}^{0.055}$&IC2391&2 \\
 48&VXR PSPC 77a&9.91&02/04&0.6000&0.5000&0.4184&0.050$_{0.045}^{0.055}$&IC2391&2 \\
 49&Cl* NGC 2682 SAND 785&14.8&02/05&0.8315&0.6500&0.3233&4.000$_{3.800}^{4.300}$&M67&3 \\
 50&Cl* NGC 2682 SAND 1477&14.6&02/05&0.8497&0.6700&0.2981&4.000$_{3.800}^{4.300}$&M67&3 \\
 \enddata
\tablenotetext{a}{1: Barnes, Sofia $\&$ Prosser (1999); 2: Patten $\&$ Simon (1996); 3: Giampapa et al. (2006) }

\end{deluxetable}

\begin{deluxetable}{ccccccccccc}
\tabletypesize{\scriptsize}
%\rotate
\tablecaption{The V-I, B-V, S$\rm_{HK}$, v$\rm_{r}$, full space motions and age for 27 wide binaries\label{tbl-3}}
\setlength{\tabcolsep}{0.02in}
\tablewidth{0pt}
\tablehead{
\colhead{ID} & \colhead{V-I}&\colhead{B-V}& \colhead{S$\rm_{HK}$}& \colhead{v$\rm_{r}$} & \colhead{$\pi$} & \colhead{U} & \colhead{V}& \colhead{W}&\colhead{[Fe/H]}&\colhead{age}\\
&&&&(km s$^{-1})$&(mas)&(km s$^{-1}$)&(km s$^{-1}$)&(km s$^{-1}$)&&(Gyr) \\
(1)&(2)&(3)&(4)&(5)&(6)&(7)&(8)&(9)&(10)&(11) \\
}
\startdata
1a&0.5210&0.8230&0.2303&76.1$\pm$7.3&5.89$\pm$2.77&-329.0&-242.5&-101.4&-0.48$\pm$0.02& \\
2a&0.6595&0.5660&0.2843&16.1$\pm$4.3&9.06$\pm$2.68&-42.5&-38.6&11.7&-0.17$\pm$0.07& \\
2b&1.0701&0.9111&0.2413&114.3$\pm$2.1&9.06$\pm$2.68&10.3 &-109.3 &54.9 &-0.40$\pm$0.08& \\
3& 0.6624&0.5699&0.2891&58.8$\pm$9.7&18.70$\pm$0.80&26.4&-65.8&-39.0&-0.25$\pm$0.02& \\
4&0.6800&0.5230&0.2689&28.3$\pm$5.4&62.95&-10.3&16.5&-19.1&-0.34$\pm$0.06& \\
5&0.6901&0.6762&0.2666&-54.3$\pm$6.1&28.3&-51.2&44.0&-18.6&0.10$\pm$0.01& \\
6&0.7420&0.5970&0.4582&-20.8$\pm$6.9&83.3&-18.5&18.6&-7.1&-0.25$\pm$0.15& \\
7&0.7754&0.6422&0.3150&-7.7$\pm$4.5&56.7&0.4&4.4&0.4&0.06$\pm$0.04&1.0$_{0.9}^{1.2}$\\
8&0.7892&0.6963&0.2856&28.5$\pm$4.9&26.76$\pm1.09$&-59.4&-8.6&30.9&-0.16$\pm$0.06&\\
9&1.0776&0.9720&0.3454&27.0$\pm$1.6&28.17$\pm0.06$&-6.4&-15.4&-26.4&-0.13$\pm$0.13&\\
10&1.1181&0.9884&0.3180&55.9$\pm$1.6&22.94$\pm1.63$&-0.3&-101.2&5.9&-0.48$\pm$0.02&\\
11&1.1316&1.0362&0.4074&69.2$\pm$5.6&29.49&35.6&-53.3&3.5&0.19$\pm$0.09&2.0$_{1.4}^{2.8}$\\
12a&1.1469&0.7750&0.3051&-37.9$\pm$1.7&198.25$\pm$0.84&-104.0&-8.5&-37.4&-0.17$\pm$0.17&5.0$_{4.0}^{6.1}$ \\
12c&2.8297&1.3938&3.0759&-54.0$\pm$1.7&198.25$\pm$0.84&-104.0&-8.5&-37.4&-0.17$\pm$0.17&5.0$_{4.0}^{6.1}$ \\
13a&1.1891&0.9822&0.3140&21.8$\pm$2.0&51.71$\pm$0.91&19.6&-9.4&28.7&-0.30$\pm$0.10& \\
13b&2.3761&1.4432&0.5585&33.8$\pm$3.1&51.71$\pm$0.91&21.6&-21.6&31.7&-0.22$\pm$0.08& \\
14a&1.4140&1.2120&0.4563&67.2$\pm$1.8&42.69&-58.5&-23.0&58.5&0.19$\pm$0.09& \\
14b&2.2270&0.6870&0.5730&60.3$\pm$6.9&42.69&-52.8&-23.5&54.5&0.05$\pm$0.09& \\
15&1.8888&1.4484&0.6202&57.7$\pm$2.0&56.92&236.9&-187.8&-229.5&-0.25$\pm$0.15&3.6$_{2.7}^{5.0}$ \\
16&1.9870&1.4550&0.74958&14.4$\pm$4.2&49.63&-28.6&-11.1&-4.2&0.15$\pm$0.05&1.2$_{0.9}^{1.4}$\\
17&2.3380&1.5710&1.0492&11.7$\pm$5.2&46.12&-33.3&3.1&11.8&-0.24$\pm$0.10&1.5$_{0.8}^{2.9}$ \\
18&2.4790&1.5020&0.7738&28.6$\pm$4.2&57.70$\pm$14.40&-22.9&-39.0&8.8&-0.02$\pm$0.03&3.5$_{2.5}^{4.5}$\\
19&2.8165&1.6318&0.9172&-53.4$\pm$4.0&40.05&48.1&-9.6&-7.5&&2.2$_{1.6}^{4.8}$\\
20& 2.8474&1.5111&2.5424&24.0$\pm$3.2&38.80&-14.0&3.4&-33.0&&\\
21& 4.2570&1.8760&1.0893&-24.6$\pm$3.9&102.00$\pm$14.00&-65.8&-3.1&40.6&&2.4$_{1.8}^{4.2}$\\
22&0.8933&0.7180&0.3031&-18.3$\pm$10.1&65.1$\pm$0.4&24.8&-24.1&15.5&0.02$\pm$0.08&1.1$_{0.9}^{1.3}$\\
23&0.7769&0.5900&0.27162&-41.9$\pm$4.2&27.5$\pm$0.5&-63.3&12.9&-17.8&0.03$\pm$0.07&1.4$_{0.4}^{2.4}$\\
24&1.8366&1.5000&1.2597&3.2$\pm$3.8&57.6&29.6&-4.0&83.1&&3.6$_{2.2}^{5.6}$\\
25& 0.8133&0.630&0.32504&-2.5$\pm$4.1&77.69$\pm$0.86&-9.5&9.7&2.1&-0.10$\pm$0.12&1.4$_{0.9}^{2.3}$ \\
26&1.5700&1.1600&0.3815&56.6$\pm$4.4&32.06$\pm$1.65&131.7&-50.6&48.2&-0.45$\pm$0.26&7.5$_{6.1}^{8.7}$ \\
27&1.6900&  &0.3835&145.0$\pm$1.8&32.3&100.3&-195.3&-46.3&-0.95$\pm$0.25&7.9$_{5.5}^{9.0}$ \\
28a&1.0215&0.8394&0.2264&-29.9$\pm$2.2&45.0&-14.0&-19.3&27.4&0.04$\pm$0.06&\\
28b&0.901&0.8470&0.2054&-32.1$\pm$5.8&45.0&-14.2&-20.4&29.3&0.03$\pm$0.07& \\
29&0.7100&0.5300&0.2189&-28.1$\pm$4.1&28.7$\pm$1.3&80.4&-38.6&-12.3&-0.40$\pm$0.02 &3.5$_{3.0}^{3.9}$\\
30&0.8800&0.9800&0.2389&44.4$\pm$1.5&9.52$\pm$1.63&101.3&-20.3&-36.3&-0.23$\pm$0.03 &1.6$_{1.0}^{2.4}$\\
31&0.3758\tablenotemark{a} &1.0400&0.1585&33.2$\pm$1.9&14.68$\pm$0.96&-19.7&-29.4&-49.3&-0.16$\pm$0.04 &2.6$_{1.3}^{3.9}$\\
32&2.0889\tablenotemark{a} &&1.0034&1.6$\pm$1.2& &&&&-0.15$\pm$0.05 &1.2$_{1.0}^{1.3}$\\
33&0.7160 &0.6600&0.1820&-61.4$\pm$12.0&19.36$\pm$1.30&-57.6&-65.3&-27.8&-0.40$\pm$0.10 &1.6$_{1.4}^{1.8}$\\
34&0.6861\tablenotemark{a} &&0.2081&28.9$\pm$1.4&&&&&-0.16$\pm$0.06 &3.4$_{2.7}^{4.2}$\\
35&0.9195\tablenotemark{a} &&0.3607&27.3$\pm$3.6&21.65&26.4&-11.3&-7.7&-0.50$\pm$0.03 &3.4$_{2.6}^{4.2}$\\
36&1.52 &1.1200&1.7876&42.7$\pm$1.3&56.02$\pm$1.21&35.8&-11.6&4.4&-0.21$\pm$0.03&4.5$_{3.6}^{5.4}$\\
\enddata
\tablenotetext{a}{V-I is from the relation in Fig.1}

%% Text for table notes should follow after the \enddata but before
%% the \end{deluxetable}. Make sure there is at least one \tablenotemark
%% in the table for each \tablenotetext.
%\tablecomments{Table \ref{tbl-1} is published in its entirety in the
%electronic edition of the {\it Astrophysical Journal}.  A portion is
%shown here for guidance regarding its form and content.}
%\tablenotetext{a}{Sample footnote for table~\ref{tbl-1} that was generated
%with the deluxetable environment}
%\tablenotetext{b}{Another sample footnote for table~\ref{tbl-1}}
\end{deluxetable}

\begin{deluxetable}{cccccccccc}
\tabletypesize{\scriptsize}
%\rotate
\tablecaption{The $T$$\rm{_{eff}}$, log $g$, mass, cooling time and reference of fifteen wide dwarfs\label{tbl-4}}
\tablewidth{0pt}
\tablehead{
\colhead{ID}&\colhead{Name} &\colhead{Sp}& \colhead{$T$$\rm{_{eff}}$}&\colhead{log $g$} &\colhead{M$\rm_{WD}$}
 &\colhead{cooling time}&\colhead{M$\rm_{i}$}&\colhead{t$\rm_{evol}$}&\colhead{Ref\tablenotemark{a}}\\
&&&(K)&&(M$_{\odot}$)&(Gyr)&(M$_{\odot}$)&(Gyr)& \\
(1)&(2)&(3)&(4)&(5)&(6)&(7)&(8)&(9)&(10)
}
\startdata
7b&LP378-537&DA&10800$\pm$120&8.21$\pm$0.05&0.732$\pm$0.032&0.6760$\pm$0.0626&3.02$\pm$0.23&0.3$_{0.2}^{0.5}$&1 \\
11b&LP786-6&DB&17566$\pm$200&7.97$\pm$0.05&0.579$\pm$0.028&0.1182$\pm$0.0135&1.56$\pm$0.29&1.9$_{1.3}^{2.7}$&2 \\
12b&40Eri B&DA&16570$\pm$350&7.86$\pm$0.05&0.540$\pm$0.019&0.1122$\pm$0.0116&1.15$\pm$0.20&4.9$_{3.9}^{6.0}$&4 \\
15b&LP791-55&DQ&6190$\pm$200&8.09$\pm$0.05&0.630$\pm$0.029&2.8244$\pm$0.6080&2.09$\pm$0.31&0.8$_{0.3}^{2.1}$&3 \\
16b&LP856-53&DA&10080$\pm$200&8.17$\pm$0.05&0.705$\pm$0.032&0.7572$\pm$0.0911&2.82$\pm$0.23&0.4$_{0.2}^{0.6}$&4 \\
17b&LP498-26&DB&16779$\pm$200&8.00$\pm$0.05&0.595$\pm$0.027&0.1475$\pm$0.0155&1.73$\pm$0.28&1.4$_{0.6}^{2.7}$&2 \\
18b&LP672-1&DA&15996$\pm$11&7.753$\pm$0.002&0.486$\pm$0.001&0.1066$\pm$0.0004&1.25$\pm$0.01&3.4$_{2.4}^{4.4}$&5 \\
19b&LP916-27&DB&10826$\pm$200&8.00$\pm$0.05&0.585$\pm$0.029&0.5365$\pm$0.0531&1.62$\pm$0.30&1.7$_{1.2}^{4.3}$&2 \\
21b&LP783-3&DZ&7590$\pm$200&8.07$\pm$0.05&0.619$\pm$0.031&1.4854$\pm$0.1856&1.93$\pm$0.31&0.9$_{0.3}^{2.7}$&3 \\
22b&L481-60&DA&10613$\pm$18&8.12$\pm$0.03&0.675$\pm$0.018&0.6167$\pm$0.0254&2.56$\pm$0.19&0.5$_{0.3}^{0.7}$&6 \\
23b&L182-61&DB&16714$\pm$200&8.07$\pm$0.05&0.605$\pm$0.029&0.1539$\pm$0.0160&1.83$\pm$0.30&1.2$_{0.4}^{2.2}$&2 \\
24b&LP888-63&DA&9408$\pm$8&7.93$\pm$0.02&0.559$\pm$0.011&0.6515$\pm$0.0159&1.35$\pm$0.11&3.0$_{1.6}^{5.0}$&3 \\
25b&CD-38 10980&DA&24276$\pm$200&8.01$\pm$0.05&0.641$\pm$0.028&0.0279$\pm$0.0043&2.21$\pm$0.29&0.7$_{0.3}^{1.6}$&6 \\
26b&LHS193B&DA&4394$\pm$200&8.10$\pm$0.05&0.632$\pm$0.032&6.6900$\pm$0.8100&2.11$\pm$0.33&0.8$_{0.3}^{1.3}$&7 \\
27b&LHS300B&DA&4705$\pm$200&7.80$\pm$0.05&0.456$\pm$0.026&4.5385$\pm$0.7298&1.25$\pm$0.01&3.4$_{2.4}^{4.4}$&7 \\
29b&G156-64&DA&7165$\pm$165&8.43$\pm$0.07&0.869$\pm$0.046&3.2983$\pm$0.4292&4.02$\pm$0.34&0.1$_{0.1}^{0.3}$&4 \\
30b&G172-4&DA&10440$\pm$240&8.02$\pm$0.07&0.613$\pm$0.043&0.5566$\pm$0.0803&1.92$\pm$0.45&1.1$_{0.3}^{1.9}$&8 \\
31b&LP592-80&DA&7520$\pm$260&8.01$\pm$0.45&0.600$\pm$0.256&1.2822$\pm$0.7289&1.78$\pm$0.20&1.3$_{0.3}^{2.0}$&1 \\
32b&G216-B14B&DA&9860$\pm$226&8.20$\pm$0.07&0.724$\pm$0.045&0.8443$\pm$0.1246&2.96$\pm$0.33&0.3$_{0.2}^{0.4}$&4 \\
33b&G116-16&DA&8750$\pm$201&8.29$\pm$0.07&0.780$\pm$0.045&1.3397$\pm$0.1785&3.37$\pm$0.33&0.2$_{0.1}^{0.3}$&4 \\
34b&G273-B1B&DA&18529$\pm$37&7.79$\pm$0.01&0.509$\pm$0.004&0.0617$\pm$0.0136&0.83$\pm$0.04&3.4$_{2.7}^{4.0}$&5 \\
35b&G94-B5B&DA&15630$\pm$65&7.89$\pm$0.01&0.555$\pm$0.005&0.1456$\pm$0.0384&1.31$\pm$0.05&3.3$_{2.4}^{4.0}$&5 \\
36b&LP358-525&DA&5620$\pm$110&8.14$\pm$0.07&0.673$\pm$0.044&4.0693$\pm$0.8147&2.54$\pm$0.46&0.5$_{0.2}^{0.8}$&4 \\

\enddata
\tablenotetext{a}{This column lists the source of $T$$\rm{_{eff}}$, log $g$. 1: Catal\'{a}n et al. 2008; 2: Voss et al. 2007; 3: Sion et al. 2009; 4: Bergeron et al. 1995; 5: Koester et al. 2009; 6: Koester et al. 2001; 7: Monteiro et al. 2006; 8: Weidemann $\&$ Koester 1984}

\end{deluxetable}

\begin{deluxetable}{cccc}
\tabletypesize{\scriptsize}
%\rotate
\tablecaption{The comparison between our ages and those from literature \label{tbl-2}}
\tablewidth{0pt}
\tablehead{
\colhead{Name} &\colhead{age$\rm_{n}$\tablenotemark{a}}& \colhead{age$\rm_{v}$\tablenotemark{b}}& \colhead{age} \\
&(Gyr)&(Gyr)&(Gyr)\\
(1)&(2)&(3)&(4) \\
  }
\startdata
 40 Eri A&$\sim$&12.2$_{8.5}^{14.5}$&5.0$_{4.0}^{6.1}$ \\
 CD-38 10983&2.5$^{7.0}_{\sim}$&2.0$_{0.4}^{3.9}$&1.2$_{0.4}^{2.2}$ \\
 CD-59 1275&5.9$_{5.4}^{6.6}$&3.7$_{3.4}^{4.7}$&1.4$_{0.4}^{2.4}$ \\
 CD-37 10500&7.4$_{1.9}^{13.0}$&4.4$_{1.4}^{7.0}$&1.1$_{0.9}^{1.3}$ \\
 \enddata
\tablenotetext{a}{ages are from Holmberg, Nordstr\"{o}m $\&$ Andersen (2009) }
\tablenotetext{b}{ages are from Valenti $\&$ Fischer (2005) }

\end{deluxetable}

%% If you use the table environment, please indicate horizontal rules using
%% \tableline, not \hline.
%% Do not put multiple tabular environments within a single table.
%% The optional \label should appear inside the \caption command.

\clearpage

%% Tables may also be prepared as separate files. See the accompanying
%% sample file table.tex for an example of an external table file.
%% To include an external file in your main document, use the \input
%% command. Uncomment the line below to include table.tex in this
%% sample file. (Note that you will need to comment out the \documentclass,
%% \begin{document}, and \end{document} commands from table.tex if you want
%% to include it in this document.)

%% \input{table}

%% The following command ends your manuscript. LaTeX will ignore any text
%% that appears after it.

\end{document}